\documentstyle[11pt,aaspp4]{article}

\lefthead{ Lee}
\righthead{Band Strength in M3}

\begin{document}

\title{CN and CH Band Strengths of Bright Giants in M3}

\author{Sang-Gak Lee}
\affil{Department of Astronomy, Seoul National University, Seoul, Korea}

\begin{abstract}
 CN and CH band strengths  for ten bright red giants in M3 have
been measured from archival spectra obtained with the MMT. A CN-CH
band strength anticorrelation is confirmed for the program stars
together  with  other stars for which there is published  
data.  This  suggests an anticorrelation between carbon  and 
nitrogen abundances with a 
constant total abundance of carbon plus nitrogen. 
However, stars which do not follow the CN-CH anticorrelation are also
found. The star III-77, which is found to have both strong 
CN and CH bands, is the most peculiar among them. While three other stars,
VZ194, VZ352, and VZ1420, which show both weak CN and CH band strengths
could be AGB stars.   
\end{abstract}


\section{Introduction}

The chemical elements C, N, O, Na, Mg, and Al vary from star to star 
within Galactic globular clusters (for reviews, see \cite{smi87}; 
\cite{sun89}, 1993; \cite{kra94}).
It has been found that in general
the nitrogen abundance is anticorrelated
with both carbon and oxygen while it is correlated with Na and Al.
These intracluster abundance inhomogeneities can be interpreted either 
by mixing of nucleosynthesized material from the deep
stellar interior during the red giant branch (RGB) phase of  evolution 
or by inhomogeneities of
primordially processed material, from which the stars were formed. 
Models of luminous cluster giants (\cite{swe79}) show that exterior to 
the main H $\rightarrow$ He burning shell are narrow zones 
in which C $\rightarrow$ N and 
O $\rightarrow$ N 
nucleosynthesis are active; the C $\rightarrow$ N zone is 
the outermost while the O $\rightarrow$ N zone
is interior to this. If material in the envelope of a cluster giant can be 
cycled through and processed in the C $\rightarrow$ N zone, then the surface 
material will be depleted in carbon and enriched in nitrogen.
Therefore the anticorrelation between surface carbon and nitrogen abundances 
among the bright giants in the clusters can be obtained.
The time taken to transport material between the base of the 
convective envelope and the C $\rightarrow$ N burning shell is 
$\sim$ 3 $\times$ $10^5$ yr
(\cite{smi92}), which is several orders of magnitude less
than the time taken to evolve up the red giant branch
($\sim$ $10^8$ yr).
If circulation reaches down into the O $\rightarrow$ N 
processing zone, depletion
of oxygen and enrichment of nitrogen result. According to 
Langer et al. (1993) Na and Al can be synthesized by proton- 
addition reactions
in the O $\rightarrow$ N burning shell, but not in 
the C $\rightarrow$ N burning shell, 
although the models employed by  
Cavallo et al. (1996, 1998) 
indicate 
that Na can be produced between these two shells.

The CN band strength is a useful indicator of nitrogen abundance
while the CH band strength provides information about the 
carbon abundance.
The nitrogen and carbon abundances derived from the CN and CH band 
strengths of bright giants in the globular clusters M3 and M13 
(\cite{bel80}; \cite{sun81}; \cite{nor84}; \cite{bro91}; 
\cite{smi96}), M5 (\cite{smi97a}), and M10
(\cite{smi97b}) have been estimated.  
The results are that giants in these clusters 
exhibit an anticorrelation 
between CN band strength and both [O/Fe] and [C/Fe] abundances,
coupled with a CN-[N/Fe] correlation, and the total [(C+N+O)/Fe]
abundance is the same among both CN-strong and CN-weak giants. 
The results support the hypothesis that C $\rightarrow$ N
and O $\rightarrow$ N processed material has been brought to the surface
of the CN-strong giants, based on the fact that
the total [(C+N+O)/Fe] is the same within the observational 
uncertainties among both CN-strong and CN-weak giants.
However, stars such as IV-59 in M5 (\cite{smi97a}) and A-I-61 in 
M10 (\cite{smi97b}), 
which do not follow the typical CN-CH band strength 
anticorrelation, are found.  The former, which shows a peculiar pattern
of abundances, is suggested to be  most likely due to 
the transfer of carbon-rich material from an unseen 
companion. The later, which
shows both weak CN and CH band strengths, is proposed as a possible
asymptotic giant branch (AGB) star.

The availability of old archival spectra of red giants in M3 
led the author to search for further stars which do
not follow the CN-CH band strength anticorrelation. 
In this study, the CN and CH 
band strengths are measured for ten red giant stars in M3.
The results are discussed along with data in the 
literature for other M3 red giant stars.

\section{Measurements of CN and CH Band Strengths}

Spectra of ten M3 giants were provided by Dr. K. Croswell. He obtained 
them with the Multiple Mirror Telescope (MMT) atop Mountain Hopkins on
two nights each in
February 1987 and March 1988.
The blue channel of the MMT spectrograph with  an 832 l/mm
grating blazed at 4300 $\AA$   in second order was used.
The resolution was 1 $\AA$  with 5 pixels per resolution element,
and the wavelength coverage ranged from 3600 $\AA$  to 4400 $\AA$.
The observed data were reduced using the standard NOVA package
at the CfA(Center for Astrophysics). The author obtained the wavelength-
calibrated data from Dr. Croswell. 

The ten stars, identified according to the star numbers from the
catalogue by  von Zeipel (1908) are listed in Table 1, together 
with $V$ and $B-V$ taken from Table III of  Cudworth (1979), 
who took these data from Sandage (1970), Johnson
and Sandage (1956), or Zhukov (1971). The absolute magnitudes 
are also listed, which are based on an apparent distance
modulus for M3 of $(m - M)_v$ =15.05 (\cite{pet93}).
Figure 1 shows the color-magnitude diagram with fiducial lines
for the AGB, HB, and RGB of M3 taken from Table 2 of Ferraro et al.(1997).
Open symbols (AGB, triangle; HB, square; RGB, circle)
are fiducial line points, filled squares are program
stars, and filled triangles are the stars from Smith et al. (1996) 
which will be discussed later.

\placefigure{fig1}

 Two stars (VZ352 and VZ1420) of the program are located in the 
region blueward of the AGB line, four stars (VZ164, VZ238, VZ265, and VZ297)
lie in the region redward of the RGB line, while the remaining four stars
(VZ194, VZ323, VZ334, and 
VZ1391) fall between these two lines. 
VZ352 and VZ1420 are likely AGB 
stars, while VZ164, VZ238, VZ265, and 
VZ297 are probably first-ascent RGB 
stars. The remaining four stars could be either AGB or RGB.
They are listed as AGB, RGB, and MID in the last column of 
Table 1.
 
Three indices which are sensitive to the strength of the CN and
CH bands in the spectra were measured. For CN indices 
the same definitions in Smith et al. (1996, 1997) are used. 
The CH index is defined as $CH(G) = -2.5 log {{\Sigma_{4270}^{4320} I_{\lambda}}
\over {{1 \over 2}(\Sigma_{4230}^{4260} I_{\lambda} + \Sigma_{4390}^{4420} I_{\lambda})}} $,
which is the same definition, excepting the continuum parts, as that in 
Smith et al. (1996, 1997). 
These three indices are denoted CN$_{(3883)}$, CN$_{(4215)}$, 
and CH(G).  An additional index, $<CN>$, which is an average of 
the CN$_{(3883)}$ index and twice the CN$_{(4215)}$ index, is used in order
to reduce the effects of measurement errors. Figure
2, a plot of the CN$_{(3883)}$ index versus the CN$_{(4215)}$ index 
shows that the 
two indices are strongly correlated in such a way that the 
CN$_{(4215)}$ index has a value of one-half the CN$_{(3883)}$ index for
the program stars.
Five measurements for each index give errors
less than 0.01 for each star in this work. 
The values of the indices obtained in this work are listed in
Table 1.

\placetable{tbl-1}
\placefigure{fig2} 

\section{The CN-CH Band Strength Anticorrelation} 

The behavior of the $<CN>$  and CH(G) indices as a function
of $M_v$ is shown in Figure 3,  and as a function
of $(B-V)$ in Figure 4. Filled squares
represent the program stars.

Four of the program stars (VZ238=AA, VZ265=IV-101, VZ297, VZ334=IV-77)
were also observed by
Smith et al. (1996). 
The difference in index values between 
Smith et al. (1996) and this study 
is assumed to consist only of a zero-point
shift since the same index definitions for CN indices
and an almost identical definition for the CH index excepting  the
continuum parts  
are adopted, although different spectra are used.
The average differences  for the above 
four stars between the S(3839), S(4142), and $M_{CH}$ indices of  
Smith et al. (1996) and the corresponding CN$_{(3883)}$, 
CN$_{(4215)}$ and CH(G)
indices of this study are 0.25, 0.36, and 0.06 respectively. 
The  filled triangles in Figures 3 
and 4 are the stars from
Smith et al. (1996), whose indices have been transformed to the
system of this study. 
The $<CN>$ index for them is obtained 
as for the program stars.  They are also plotted in 
Figure 1 and listed in the bottom ten rows of Table 1. 
One star (VZ1208) could not be included in Figures 1 and 4
due to a lack of $(B-V)$ data.

\placefigure{fig3}
\placefigure{fig4}

Two separate sequences of CN strength as a function of
$(B-V)$ were found in M3 by Suntzeff (1981). The same feature
is seen in Figure 3 of Norris and Smith (1984). A similar
distribution of $<CN>$ index is noticed in Figures
3 and 4 of this work. 
Since the CN strength of a star varies according to its position
in the color-magnitude diagram, it is assumed that the lower
envelope to the distribution of points in Figures 3 and 4 
represents the minimum value of $<CN>$ at any $M_v$ or $(B-V)$
on the giant branch.
The lines 
in Figures 3 and 4 
represent the equations
$<CN>$ = --0.11 $\times$ $M_v$ + 0.096 and $<CN>$ = 0.27
$\times$ $(B-V)$ -- 0.058. A parameter $D_{<CN>}$ is defined
as a measure of the displacement of any observed point above
these lines. The average $D_{<CN>}$ for each star is listed 
in Table 1. The average of $D_{<CN>}$ for the  eleven stars 
around the lower envelope is $0.00\pm 0.03$, while that 
for the remaining nine stars is $0.16\pm0.05$.    
The bimodal nature of the distribution is apparent 
with a separation between the averages of 0.16 which
is more than three times the dispersion for each group.
Therefore an arbitrary line of $D_{<CN>}$ $\sim$ 0.07 separates stars
into CN-weak and CN-strong groups.
A similar distribution of CN-strong and 
CN-weak giants has been found in NGC 3201 (\cite{smi82}
), NGC 6752
(\cite{nor81b}; \cite{sun91}), 
M4 (\cite{nor81a};
\cite{sun91}), and M5 (\cite{smi83}).

Among  nine CN-strong stars, four program stars 
(VZ164, VZ238=AA, VZ265, and VZ297) and 
one star (VZ1000) from Smith et al. (1996) are 
supposed to be  
red giants according to their positions in the color-
magnitude diagram.  The remaining four RGB 
stars (VZ1397, I-21, A and BF), which are all from
Smith et al.(1996), 
belong to the CN-weak group. 
The four program stars between the
AGB and RGB fiducial lines belong to either the CN-strong
(VZ334) or CN-weak group (VZ194, VZ323, and VZ1391), and
two program stars (VZ352 and VZ1420) in the AGB region are CN-weak.
However among the remaining five
stars from Smith et al. (1996) 
one (III-77) on the AGB fiducial line
and VZ1208 without $(B-V)$ data are CN-strong, whereas the three
stars (III-28, II-46, and VZ1127) are found to be CN-weak. 	 
The symbols S and W in the last column of Table 1 
indicate CN-strong and CN-weak respectively.

 The minimum value of $<CN>$ at each state on the
red giant branch may be affected by not only 
the temperature and
gravity but also the
surface abundances
of nitrogen and carbon.
This means that the same $D_{<CN>}$
does not necessarily imply the same [N/Fe] because the nitrogen and
carbon abundances
for $D_{<CN>}$ = 0 could differ from star to star either
on the RGB or on the AGB.
The synthetic CN indices  plotted in Figures 3 (RGB stars) and
4 (AGB stars) of Briley and Smith (1993) indicate that the minimum value
of their S(3839) indices on the RGB and the AGB differ according to
the nitrogen abundance as well as the carbon abundance. This was
called the "weak CN effect" coupled with the "weak G-band effect" by
Suntzeff and Smith (1991).
The nitrogen abundance data for
seven stars from Smith et al (1996) give a better linear correlation with
$<CN>$ than with $D_{<CN>}$. Even though this may be
due to the fact that these stars are
all within a narrow range of $M_v$ and $(B-V)$,
the index $<CN>$ is preferred for use in further
discussion.

For the CH(G) index, there is a tendency
for it to increase in value up to $(B-V)$
$\sim$ 1.0 and $M_v$ $\sim$ --1.3 and then decrease as a star 
becomes brighter and redder. A similar distribution is found among the stars
in M10 (\cite{bri93}). 
The synthetic G-band index shows a tendency which is different for 
the RGB stars and the AGB stars. 
But it is not easy to define the
relation of the CH(G) index with either $(B-V)$ or $M_V$ 
separately for the RGB and the AGB 
with the small sample of stars in this study. 
However the CH(G) index is found to be linearly correlated with 
[C/Fe] for the seven stars of Smith et al (1996), and so we use the 
CH(G) index  itself for further discussion.

\placefigure{fig5}

A plot of 
the $<CN>$ index versus the CH(G) index in Figure 5 shows that
there is an anticorrelation between $<CN>$ and CH(G) 
for most CN-strong and CN-weak stars; however some
stars, III-77 (CN-strong), and VZ194, VZ352, VZ1391, and
VZ1420 (CN-weak),  
do not follow that relation. However if $D_{<CN>}$ is used instead of
$<CN>$, VZ1391 could be included among the stars 
which follow the CN-CH band strength anticorrelation. Therefore it will be
excluded from further discussion.

III-77 may be an AGB star because its position
in the color-magnitude diagram is right on the AGB
fiducial line. Unfortunately, Smith et al. (1996)
did not obtain [N/Fe] and [C/Fe] for this star. If linear
relations between [N/Fe] abundance and the $<CN>$ index, 
and [C/Fe] abundance and the CH(G) index, which are derived 
from the seven stars of Smith et al. (1996), 
are applied to III-77,
the nitrogen and carbon abundances of this star are 
found to be [N/Fe]
$\sim$ 1.00 and [C/Fe] $\sim$ 0.02. The nitrogen 
enhancement can be explained by the mixing process
during the previous RGB evolutionary state 
but the undepleted carbon abundance is hard 
to understand if  mixing has occurred.  Although 
its CH(G) index is outside of the linear transformation
limits, which  makes the carbon abundance quite uncertain, it 
still seems acceptable that carbon in this star
is not as depleted as in other AGB stars.
III-77 could be either a 
star without mixing whose nitrogen has been enriched 
by processed material transfered from a companion
or  a star having  entirely different primordial
abundances of C and N.

Among the remaining three stars which do not follow the CN-CH
band strength anticorrelation,  
the positions of  VZ352 and VZ1420 are found to be blueward 
of the fiducial AGB line in the color-magnitude diagram 
indicating that they are AGB stars  while VZ194 
could be either AGB or RGB, although its position in Figure 1 gives more
possibility of AGB than of RGB.  
Their $<CN>$ and  
CH(G) indices are all outside of the linear
transformation limits. They are all bluer than 
$B-V$ = 1.0 and fainter than $M_v$ $\sim$ --1.3. 
Therefore the apparent weakness 
of the CN and CH band strengths may in part be due to the 
effects of temperature and gravity. But these
effects may not be entirely dominant 
because VZ323, a star located in a similar part of
the color-magnitude diagram  with $M_v$ = --0.37
and $B-V$ = 0.83 does belong to the CN-strong group.  
If linear transformations are extrapolated to
these stars, they give [C/Fe] = --1.80, --2.00, and --1.30, 
with [N/Fe] = 0.13, 0.16, and 0.05, for
VZ352, VZ1420, and VZ194 respectively. 
However it has been shown that the same 
CN band strength would give 
significantly different nitrogen abundances  
between the RGB stars and the AGB stars,
seen Figures 3 and 4 of 
Briley and Smith (1993). 
This is due to the fact that, 
as the carbon abundance decreases less than [C/A] $\sim$ -1.0,
the formation of CN molecules drops significantly regardless of 
the value of the nitrogen abundance. There is thus a
"weak CN effect" coupled with a "weak G-band effect"
on the AGB, as  noted by Suntzeff and Smith (1991). 
Therefore, if not only VZ352 and VZ1420 but also 
VZ194 are AGB stars, the low values of their  
$<CN>$ indices could imply significantly more abundant 
nitrogen than is indicated by the linear transformation.

\section{Discussion and Summary}

The CN-CH band strength anticorrelation among bright giants of 
a number of metal-poor globular clusters (\cite{kra94}) 
has been used as one item of evidence for the  
evolutionary scenario 
which postulates the production  
of the processed material, nitrogen, at the expense of
carbon and oxygen in the CNO burning shell and 
subsequent transport to the stellar surface (\cite{swe79}).
The fact that the sum of the oxygen,
carbon, and nitrogen abundances is approximately the same
in giant stars for which the three elements have been observed
 (\cite{bro91}; \cite{smi96}),
 and other anticorrelations of O-N and O-Na which could also be 
accounted for by a deep mixing scenario (\cite{lan93}; 
\cite{cav96}, 1998), except the Mg-Al anticorrelation which 
poses some difficultes to an entirely deep mixing
origin, suggest that a considerable amount of the  
abundance variations of C, N, O, Na, and Al is contributed by
mixing during the RGB evolutionary stage. 

However the CN-CH anticorrelation and the bimodal 
CN band distribution have been found to exist among not only
evolved stars (\cite{nor79}, 1982; 
\cite{smi89};
\cite{bri89}) but also the unevolved stars in 
NGC 6752 (\cite{sun89}) and 47 Tuc (\cite{bri94}
). This suggests that at least some component of
the C and N abundance inhomogeneities observed among the 
red giants in these clusters were established before the
red giant branch.     

Therefore the confirmation of a CN-CH anticorrelation and 
the bimodal distribution
of CN bands among the red giant stars in M3  
may not give any decisive clues for 
star-to-star variations of light elements in this cluster.
However the stars which do not follow the anticorrelation,
are worth discussing.

The star III-77, which is the weirdest  among them,
is found to be as nitrogen enriched as  
other CN-strong stars but not depleted in carbon at all. 
This star is similar to IV-59 of M5 (\cite{smi97a})
in the sense that both
stars violate the CN-CH anticorrelation, 
but IV-59 in M5 has been found to violate the O-N 
anticorrelation as well.
Their peculiar abundance pattern could originate either from 
the transfer of processed material from an unseen
companion or from inhomogeneity of the primordial material.
As Smith et al. (1997) speculate IV-59 to be a former CH star,
so do we view III-77 as a cluster CH star. It may be related to 
the very metal-deficient field CH stars, CS 22892-52 
(\cite{sne94}, 1996; \cite{mcw95};
\cite{cow95}), and CS 22948-27 and CS 29497-34 
(\cite{bar97}), which show overabundances of 
C and N with respect to iron. But III-77 is less metal-deficient. 

The classical CH stars (\cite{kee42}) are Population II giants
whose spectra show not only enhanced CH absorption but also
strong Swan bands of $C_2$. They are found to have enhanced
abundances of carbon and s-process elements. A handful of 
CH stars are found in the globular clusters $\omega$ Cen 
(\cite{har62}; \cite{dic72}; \cite{mcc77}), M22 
(\cite{hes77}; \cite{mcc77}; \cite{hes79}), M55 
(\cite{smi82}), M2 (\cite{zin81}),
and M14 (\cite{cot97}). However the fact that most globular cluster CH
stars are found not to have enhanced abundances of 
s-process elements leads to the conclusion that  
their anomalous carbon abundances 
or abnormally strong CH absorptions are due to 
incomplete CN processing (\cite{van92}).
The star III-106 of M22, which was identified on the basis of
DDO photometry (\cite{hes77}) and on the basis of a
low-resolution spectrum (\cite{mcc77}) as a CH star,
has subsequently been found not to be a  genuine CH star (\cite{van92}
). So far, among the CH-enhanced stars in globular clusters, 
RGO 55 (\cite{har62}) and RGO 70 
(\cite{dic72}) in $\omega$ Cen,
and a CH star in M14 have been found to be 
genuine CH stars (\cite{cot97}). 

The classical CH stars have halo characteristics in their abundances
and kinematics. Therefore the origin of the enhanced abundances 
of carbon and s-process elements has been a puzzle, since stars of
0.8 $M_{\sun}$  may not experience the third dredge-up mechanism
during their ascent of the asymptotic giant branch (\cite{ibe75}). Since all
field CH stars are found to be binaries (\cite{mcc90}), 
the peculiar abundances
can be explained by mass transfer. The binaries are composed of
a red giant primary and a white dwarf secondary (\cite{mcc90}
). The mass is transferred  via a stellar wind 
or Roche lobe overflow during the ascent of the white dwarf 
progenitor up the AGB. 

However Gunn and Griffin (1979) have shown that the star III-77 
does not exhibit any of the variations in radial velocity expected 
were it a binary 
and also have concluded that there is a 
deficiency of binaries in the globular cluster M3. Therefore
if the pecularity of III-77 comes from the binary nature of a CH star,
it would have important implications not only for nucleogenesis
and globular cluster abundance anomalies but also for the formation,
evolution, and destruction of binaries in dense environments.
If III-77 had been enriched  in C and N at its formation, 
it would be 
expected to have enrichments in heavy elements  as well, 
such as calcium and iron.
This star surely requires a high resolution spectroscopic study
to solve all these puzzles.

In the case of the other three stars in M3 which do not show a 
CN-CH anticorrelation, they are similar to A-I-61
in M10 (\cite{smi97b}). 
VZ352 and VZ1420 which 
could be AGB stars,  have carbon depletions and nitrogen levels 
nearly normal and  VZ194, 
whose evolutionary status is a little uncertain, has less carbon 
depletion and  
nitrogen in  slightly less abundance  than VZ352 and VZ1420.
As mentioned in the previous section, the weakness of CN bands
in AGB stars is due to a "weak CN effect" coupled with the
"weak G-band effect" on the AGB. If the star VZ194 is an
AGB star too, then its abnormally weak CN and CH bands 
can be explained by these effects. 
The position of VZ194 in the CM
diagram gives a high possibility of it being an AGB star.

In summary the CN-CH band strength anticorrelation and the bimodal
distribution of CN bands among the red giant stars in M3 
are confirmed and stars which do not follow
the CN-CH anticorrelation are found.
Among them, III-77 is the most peculiar and shows strong
CN and CH bands. It may be explained by the binary
nature of cluster CH stars, but the fact that the chance of
finding binary stars in M3 seems quite low makes this
scenario less certain. It could also be explained by the 
inhomogeneous enrichment of
primordial material, but in that case we would expect  
other heavy element enrichments too.  
The weaknesses of the CH and CN bands in VZ194,
VZ352, and VZ1420 could be explained by the "weak CN effect"
coupled with the "weak G-band effect" of AGB stars.

\acknowledgments

The author would like to thank Dr. K. Croswell for providing 
the spectra used in this study. She is grateful to the 
anonymous referee, whose comments and English corrections 
allowed her to appreciably improve the paper. 
This work was supported by
the Korean Ministry of Education through grant BSRI-98-5411.

\begin{table}
\dummytable\label{tbl-1}
\end{table}

\clearpage

\figcaption[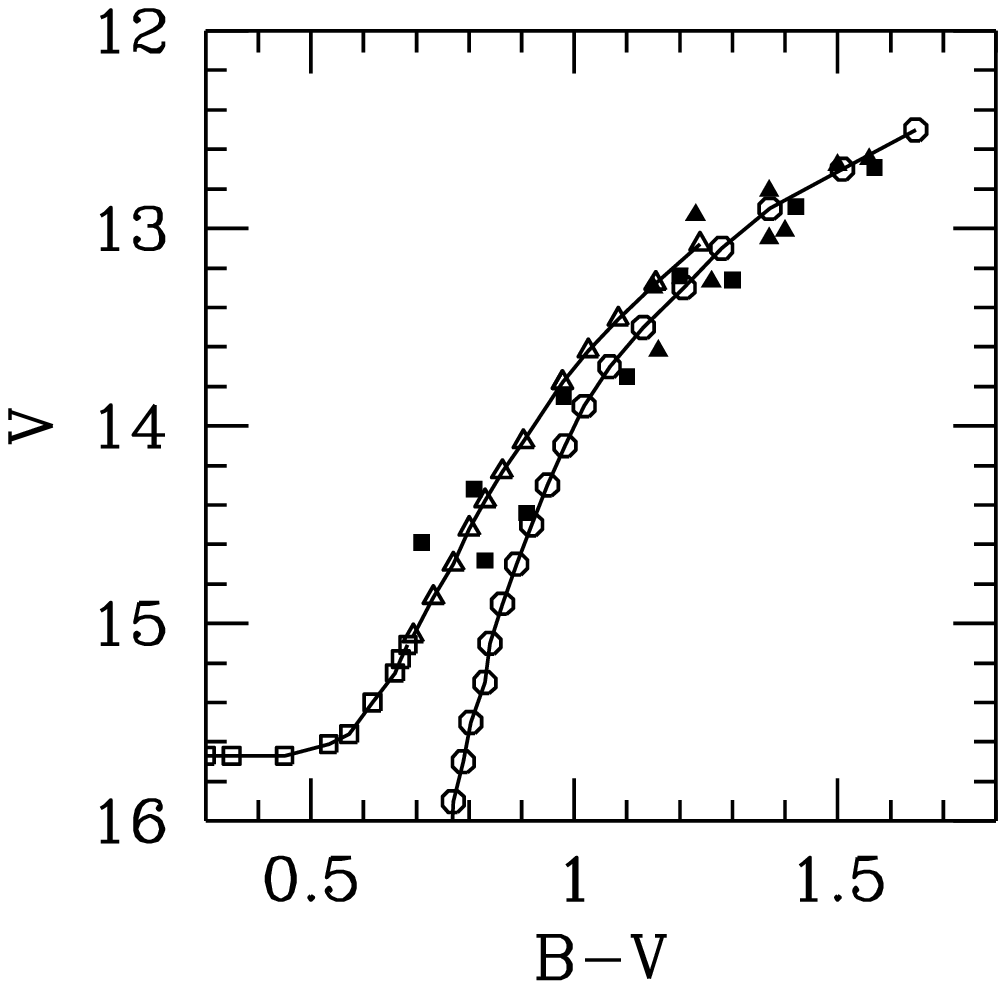]{$V$ versus $(B-V)$ color-magnitude diagram
of the globular cluster M3. Fiducial lines for the 
AGB, RGB, and HB (open triangles, open 
circles, and open squares respectively)  are taken from
Ferraro et al. (1997).  Filled squares are program stars and filled
triangles are stars from Smith et al (1996). \label{fig1}} 

\figcaption[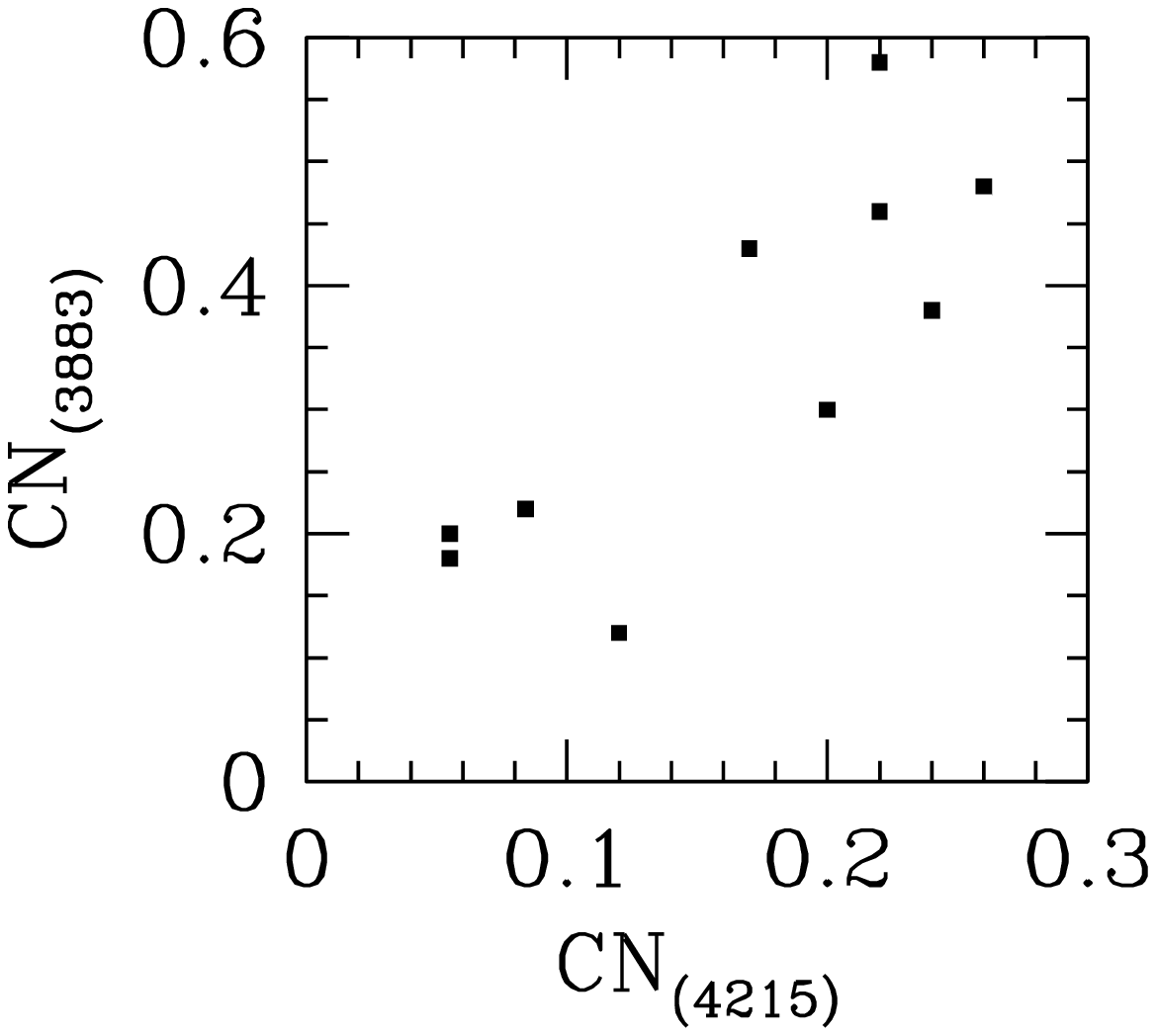]{A plot of the CN$_{(3883)}$ index versus 
the CN$_{(4215)}$ index for the program stars. The two indices 
are strongly correlated such that the CN$_{(3883)}$ index 
has a value twice that of the CN$_{(4215)}$ index. \label{fig2}}

\figcaption[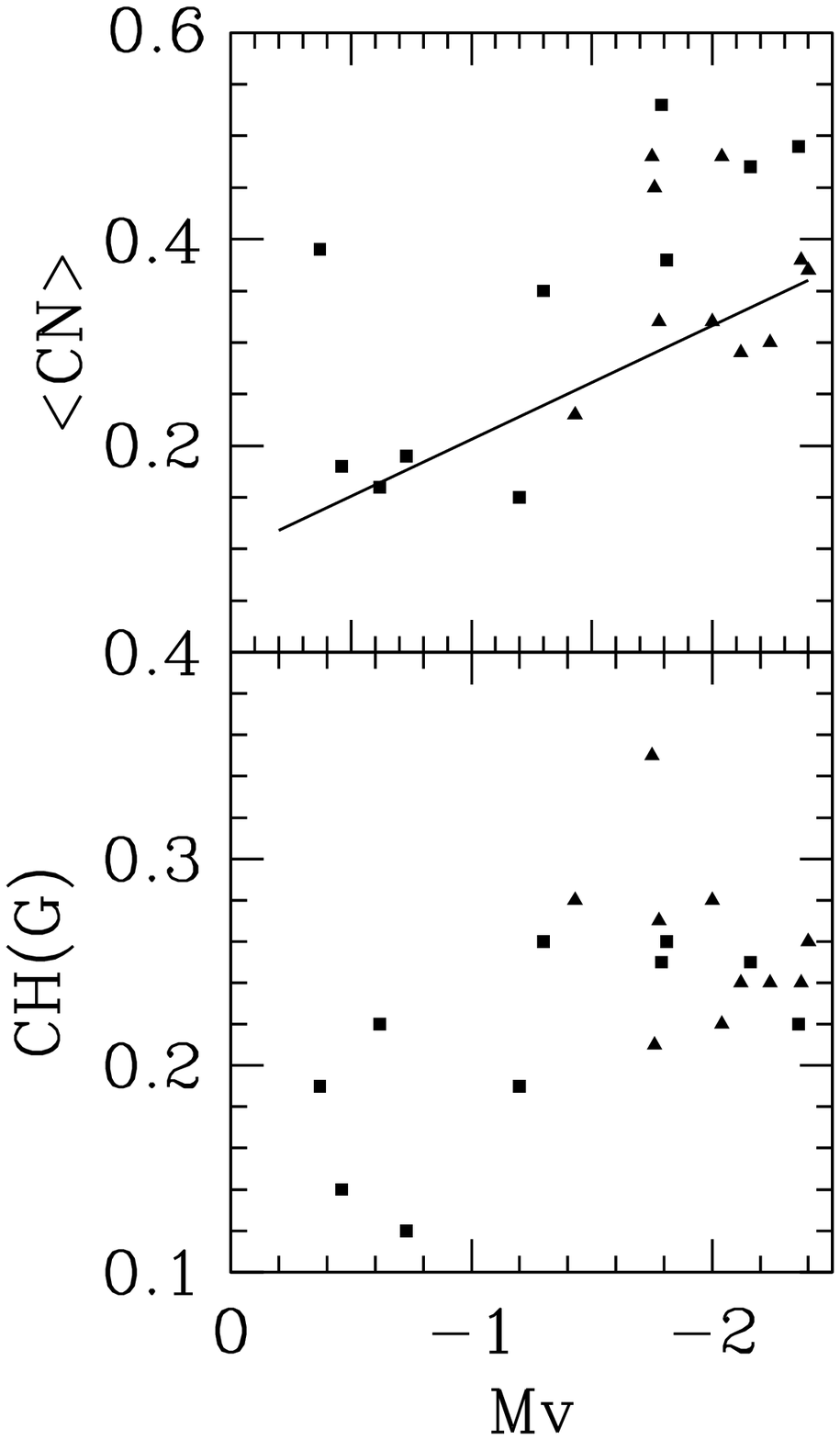]{$<CN>$ and CH(G) are plotted as 
a function of absolute magnitude for M3 giant stars.  
The mean line of $<CN>$ = --0.11 $\times$ 
$M_v$ + 0.096 for CN-weak stars 
is drawn. Symbols used are the same as for Figure 1.  \label{fig3}}

\figcaption[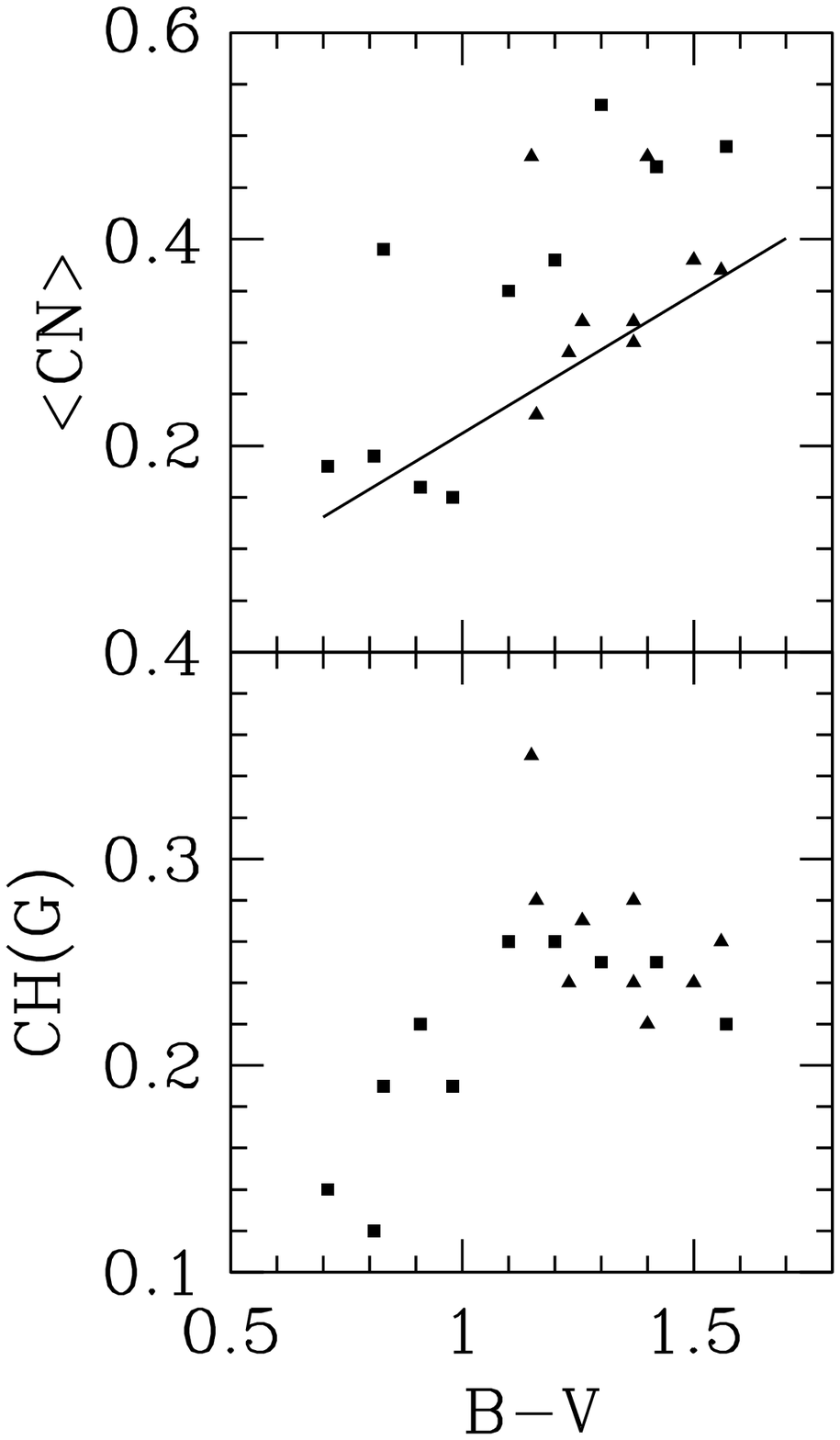]{$<CN>$ and CH(G) are plotted as
a function of $(B-V)$ color for M3 giant stars. 
The mean line of $<CN>$ = 0.27 $\times$ $(B-V)$ --
0.058 for CN-weak stars is drawn. Symbols used are the 
same as for Figure 1. \label{fig4}}

\figcaption[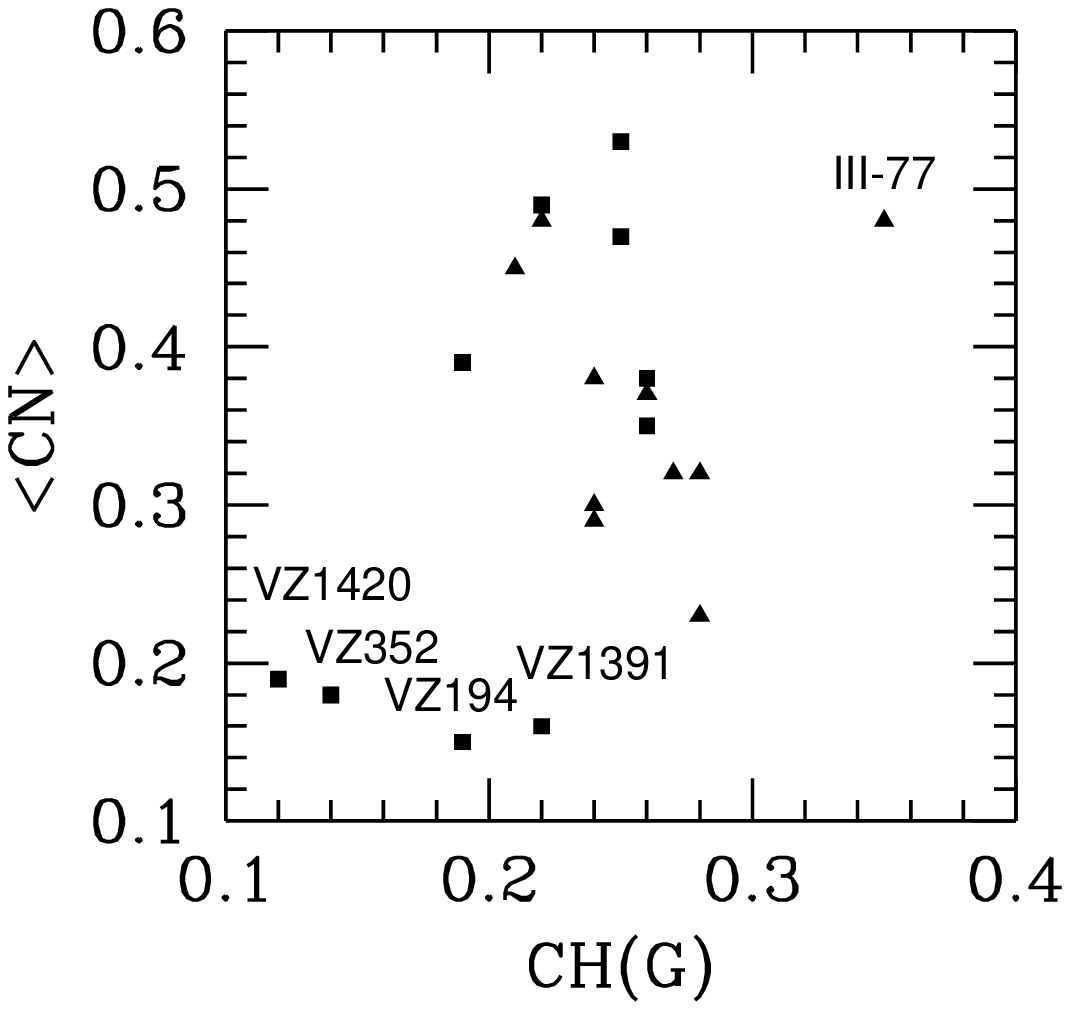]{A plot of the $<CN>$ index versus 
the CH(G) index.
A CN-CH band strength anticorrelation is shown for most 
stars, but  not for III-77 and four CN-weak stars. VZ 1397
is included in the stars which follow the $D_{<CN>}$-
CH(G) anticorrelation, therefore it is excluded from further
discussions.
Symbols used are the same as for Figure 1. \label{fig5}}


\clearpage

\begin{thebibliography}{}

\bibitem[Barbuy et al. 1997]{bar97} Barbuy, B., Cayrel, R., Spite, M.,
    Beers, T. C., Spite, F., Nordstrom, B., and Nissen, P. E. 1997, \aap, 317, L63
\bibitem[Bell and Dickens 1980]{bel80} Bell, R. A., and Dickens, R. J. 1980,
    \apj, 242, 657
\bibitem[Briley et al. 1989]{bri89} Briley, M. M., Bell, R. A., Smith, G. H., and 
    Hesser, J. E. 1989, \apj, 341, 800
\bibitem[Briley et al. 1994]{bri94} Briley, M. M., Hesser, J. E., 
    Bell, M., and Smith, G. H. 1994, \aj, 108, 2183
\bibitem[Briley and Smith 1993]{bri93} Briley, M. M., and Smith, G. H. 1993,
    ASP Conf. Ser. 48, The Globular Cluster-Galaxy Connection, edited
    by G. H. Smith and J. P. Brodie (ASP, San Francisco), p. 184
\bibitem[Brown et al. 1991]{bro91} Brown, J. A., Wallerstein, G. and Oke, J. B.
    1991, \aj, 101, 1693   
\bibitem[Cavallo et al. 1996]{cav96} Cavallo, R. M., Sweigart, A. V.,
    and Bell, R. A. 1996, \apj, 464, L79
\bibitem[Cavallo et al. 1998]{cav98} Cavallo, R. M., Sweigart, A. V.,
    and Bell, R. A. 1998, \apj, 492, 575

\bibitem[Cote et al. 1997]{cot97} Cote, P., Hanes, D. A., McLaughlin, D. E.,
    Bridges, T. J., Hesser, J. E., and Harris, G. L. H. 1997, \apj, 476, L15

\bibitem[Cowan et al. 1995]{cow95} Cowan, J. J., Thielemann, F. K., 
    and Truran, J. W. 1995, \aapr, 29, 447
\bibitem[Cudworth 1979]{cud79} Cudworth, K. M. 1979, \aj,
    84, 1312
\bibitem[Dickens 1972]{dic72} Dickens, R. L. 1972, \mnras, 159, 7P
\bibitem[Ferraro et al. 1997]{fer97} Ferraro, F. R., Carretta,
    C. E., Corsi, C. E., Fusi Pecci, F., Cacciari, C., 
    Buonanno, R., Paltrinieri, B., and Hamilton, D. 1997, \aap, 
    320, 757
\bibitem[Gunn and Griffin 1979]{gun79} Gunn, J. E., and Griffin, R. F. 
    1979, \aj, 84, 752
\bibitem[Harding 1962]{har62} Harding, G. A. 1962, Observatory, 82, 205
\bibitem[Hesser and Harris 1979]{hes79} Hesser, J. E., and Harris, G. L. H.
    1979, \apj, 234, 513
\bibitem[Hesser et al. 1977]{hes77} Hesser, J. E., Hartwick, F. D. A., and
    McClure, R. D. 1977, \apjs, 33, 471
\bibitem[Iben 1975]{ibe75} Iben, I. 1975, \apj, 196, 525
\bibitem[Johnson and Snadage 1956]{jon56} Johnson, H. L., and 
    Sandage, A. R. 1956, \apj, 124, 379
\bibitem[Keenan 1942]{kee42} Keenan, P. C. 1942, \apj, 96, 101
\bibitem[Kraft 1994]{kra94} Kraft, R. P. 1994, \pasp, 106, 553
\bibitem[Langer et al. 1993]{lan93} Langer, G. E., Hoffman, R.,
    and Sneden, C. 1993, \pasp, 105, 301
\bibitem[McClure and Norris 1977]{mcc77} McClure, R. D., and Norris, J.
    1977, \apj, 216, L101
\bibitem[McClure and Woodworth 1990]{mcc90} McClure, R. D., and Woodworth,
    A. W. 1990, \apj, 352, 709
\bibitem[McWilliam et al. 1995]{mcw95} McWilliam, A., Preston, G. W., Sneden,
    C., and Schectman, S. 1995, \aj, 109, 2736
\bibitem[Norris 1981]{nor81a} Norris, J. 1981, \apj, 248, 177
\bibitem[Norris et al. 1981]{nor81b} Norris, J., Cottrell, P. L., Freeman,
    K. C., and Da Costa, G. S. 1981, \apj, 244, 205
\bibitem[Norris and Freeman 1979]{nor79} Norris, J., and Freeman, K. C. 
    1979, \apj, 230, L179
\bibitem[Norris and Freeman 1982]{nor82} Norris, J., and Freeman, K. C.
    1982, \apj, 254, 143
\bibitem[Norris and Smith 1984]{nor84} Norris, J., and Smith, G. H. 1984,
    \apj, 287, 255
\bibitem[Peterson 1993]{pet93} Peterson, C. 1993, ASP Conf.
    Ser. 50, Structure and Dynamics of Globular Clusters,
    ed S. Djorgowski and G. Meylan (San Francisco: ASP), 337
\bibitem[Sandage 1970]{san70} Sandage, A. R. 1970, \apj, 162, 841

\bibitem[Sneden et al. 1994]{sne94} Sneden, C., Preston, G. W., McWilliam,
    A., and Searle, L. 1994, \apj, 431, L27
\bibitem[Sneden et al. 1996]{sne96} Sneden, C., McWilliam, A., Preston,
    G. W., Cowan, J. J. Burris, D., and Armosky, B. J. 1996 \apj, 468, 819
\bibitem[Smith 1987]{smi87} Smith, G. H. 1987, \pasp, 99, 67
\bibitem[Smith et al. 1989]{smi89} Smith, G. H., Bell, R. A., and
    Hesser, J. E. 1989, \apj, 341, 190
\bibitem[Smith and Fulbright 1997]{smi97b} Smith, G. H., and Fulbright,
    J. P. 1997, \pasp, 109, 1246

\bibitem[Smith and Norris 1982]{smi82} Smith, G. H., and Norris, J. 
    1982, \apj, 254, 149
\bibitem[Smith and Norris 1983]{smi83} Smith, G. H., and Norris, J.
    1983, \apj, 264, 215
\bibitem[Smith and Tout 1992]{smi92} Smith, G. H., and Tout, C. A.
    1992, \mnras, 256, 449
\bibitem[Smith et al. 1996]{smi96} Smith, G. H., Shetrone, M. D.,
    Bell, R. A., Churchill, C. W., and Briley, M. M. 1996, 
    \aj, 112, 1511
\bibitem[Smith et al. 1997]{smi97a} Smith, G. H., Shetrone, M. D.,
    Briley, M. M., Churchill, C. W., and Bell, R. A. 1997, 
    \pasp, 109, 236
\bibitem[Suntzeff 1981]{sun81} Suntzeff, N. B. 1981, \apjs,
    47, 1
\bibitem[Suntzeff 1989]{sun89} Suntzeff, N. B. 1989, in Abundance Spread Within
    Globular Clusters; Spectroscopy of Individual Stars, IAU Gen. Assembly No.  
    20, edited by G. Cayrel de Strobel, M. Spite, and T. Lloyd Evans (Observatoire
    de Paris, Paris), p. 71
\bibitem[Suntzeff 1993]{sun93} Suntseff, N. B. 1993, APS Conf.
    Ser. 48,  The Globular Cluster-Galaxy
    Connection, edited by G. H. Smith and J. P. Brodie (ASP, San Francisco),
    p. 167
\bibitem[Suntzeff and Smith 1991]{sun91} Suntzeff, N. B., and 
    Smith, V. V. 1991, \apj, 381, 160
\bibitem[Sweigart and Mengel 1979]{swe79} Sweigart, A. V., and 
    Mengel, J. G. 1979, \apj, 229, 624
\bibitem[Vanture and Wallerstein 1992]{van92} Vanture, A. D., 
    and Wallerstein, G. 1992, \pasp, 104, 888 

\bibitem[von Zeipel 1908]{zei08} von Zeipel, M. H. 1908, 
    Ann. Obs. Paris 25, F1
\bibitem[Zhukov 1971]{zhu71} Zhukov, L. V. 1971, Trudy Pulkova Obs. Ser.
    2, 78, 160
\bibitem[Zinn 1981]{zin81} Zinn, R. 1981, \apj, 251, 52

\end{thebibliography}
\end{document}